\documentclass[conference]{IEEEtran}
\IEEEoverridecommandlockouts
\usepackage{graphicx}
\usepackage{subcaption}
\usepackage{cite}
\usepackage{multirow}
\usepackage{booktabs}
\usepackage{amsmath,amssymb,amsfonts}
\usepackage{algorithmic}
\usepackage{graphicx}
\usepackage{textcomp}
\usepackage{xcolor}
\usepackage{url}
\def\BibTeX{{\rm B\kern-.05em{\sc i\kern-.025em b}\kern-.08em
    T\kern-.1667em\lower.7ex\hbox{E}\kern-.125emX}}
\begin{document}

\title{On the Robustness of Audio Deepfake Detection under Audio Watermarking\\}

\author{
\IEEEauthorblockN{
Zi Qian Yong\textsuperscript{1},
Ajinkya Kulkarni\textsuperscript{2},
Julia K. Lau\textsuperscript{1},
Hwa Hui Tew\textsuperscript{1},
Shu-Min Leong\textsuperscript{1}, \\
Rapha\"{e}l C.-W. Phan\textsuperscript{1},
Sébastien Marcel\textsuperscript{2}
}

\IEEEauthorblockA{\textsuperscript{1}School of Information Technology, Monash University, Malaysia campus}

\IEEEauthorblockA{\textsuperscript{2}Idiap Research Institute, Switzerland}


\IEEEauthorblockA{
\{zi.yong, julia.lau, hwa.tew, leong.shumin, raphael.phan\}@monash.edu\\
\{ajinkya.kulkarni, marcel\}@idiap.ch
}
}

\maketitle

\begin{abstract}
Recent advances in generative audio models have enabled highly realistic synthetic speech, increasing the importance of reliable audio deepfake detection (ADD) systems. While prior studies have primarily focused on adversarially optimized perturbations, the robustness of ADD systems under realistic signal transformations remains insufficiently understood. 
In this work, we investigate the impact of audio watermarking on ADD systems by treating watermarking as a structured, non-adversarial perturbation rather than a conventional attack mechanism. Using a watermark-based evaluation framework built upon WavMark, we evaluate multiple self-supervised learning (SSL), Convolutional Neural Network (CNN) and Graph Neural Netrowk (GNN)-based ADD models across several benchmark datasets. Beyond conventional detection metrics, we further analyze watermark-induced representation shifts using Fréchet Distance, cosine similarity, and L2 distance in the embedding space. Experimental results reveal a strong dataset-dependent behavior: watermarking causes substantial performance degradation on ASVspoof 2021 LA and DF, while exhibiting limited impact on ASVspoof 2024, FoR, and ITW. Moreover, large embedding-space shifts are strongly associated with severe detection degradation, suggesting that watermark-induced perturbations can substantially alter the feature representations relied upon by current ADD systems. These findings demonstrate that benign signal transformations designed for content protection can expose previously overlooked robustness vulnerabilities in audio deepfake detection systems. Our code is available at \url{https://github.com/ziqian0925/wm-ADD-robustness.git}
\end{abstract}

\begin{IEEEkeywords}
Audio deepfake detection, digital watermarking, adversarial robustness, distribution shift, feature perturbation.
\end{IEEEkeywords}

\section{Introduction}
Advancements in generative models, particularly Generative Neural Networks (GANs) 
and diffusion models, have enabled the generation of realistic synthetic speech and music that is increasingly difficult to distinguish from real audios \cite{li2023freevc}\cite{wang2023neural}. Despite their ability to capture intricate audio patterns with high fidelity \cite{baoueb2024specdiff}\cite{donahue2018adversarial}, concerns regarding authenticity and security remain critical for these models. For example, in March 2019, scammers used AI-generated audio to impersonate a CEO, resulting in a British energy firm being defrauded of \$243,000 \cite{forbesVoiceDeepfake}. To alleviate the risk of audio impersonation fraud, researchers have developed audio deepfake detection (ADD) systems to distinguish between real (bonafide) and fake (spoof) audio \cite{truong2024temporal} \cite{kheir2025bicrossmamba}\cite{liu2025nes2net}. Apart from that, commercial solutions such as Modulate \cite{modulate2024} reflect growing industry interest in addressing audio authenticity and security. 

Existing studies on adversarial robustness primarily focus on optimized perturbations designed to manipulate model predictions under carefully defined attack objectives.
For example, Rabhi et al. \cite{rabhi2024audio} proposed GAN-based adversarial attacks that generate synthetic histograms specifically optimized to fool the Deep4SNet audio deepfake detector. Similarly, Farooq et al. \cite{farooq2025transferable} proposed a transferable GAN-based adversarial attack framework that leverages an ensemble of surrogate ADD models and transcription-preserving objectives to generate high-quality adversarial perturbations capable of bypassing the ADD models. Although these attacks reveal important security vulnerabilities, they do not 
account for the effects of non-adversarial structured signal modifications introduced by audio authenticity techniques such as audio watermarking. This raises an important question: can audio watermarking expose hidden robustness vulnerabilities in current ADD systems?

Audio watermarking and ADD systems both support media authenticity and security from different perspectives. Watermarking provides content provenance and ownership verification, whereas ADD aims to detect manipulated or synthetic audio. However, the interaction between these two techniques remains largely unexplored. 
In particular, it is unclear whether watermark-induced signal modifications can alter ADD feature representations and degrade detection performance. Hence, in this work, we investigate how the signal changes introduced by audio watermarking influence the feature representations and decision behavior of ADD models under a black-box setting. Specifically, we compare the detection performance and embedding-space characteristics before and after watermark embedding using representation-level analyses such as Fréchet Distance, cosine distance and L2 distance.

To the best of our knowledge, prior studies have not 
investigated how signal modifications introduced by watermark affect the robustness and representation stability of ADD systems across multiple architectures and datasets. Our key contributions are summarized as follows:
\begin{itemize}
  \item We investigate audio watermarking as a structured signal perturbation framework to evaluate the robustness of audio deepfake detection (ADD) systems.
  \item We conduct extensive evaluations across different state-of-the-art ADD architectures on five datasets, revealing that the impact of watermark perturbation is highly dataset-dependent, with substantial robustness degradation observed on datasets ASVspoof 2021 LA and DF.
 \item We introduce a representation-level analysis using Fréchet Distance, cosine similarity and L2 distance to measure how watermarking changes the embedding space and affects the robustness of ADD systems, revealing that larger embedding-space shifts are strongly associated with severe detection degradation, and thus show that watermark perturbations can substantially alter latent feature representations and influence the detection behavior.
\end{itemize}

\section{Literature Review}
\subsection{Adversarial Attacks on ADD}
Recent studies have demonstrated that ADD systems are highly vulnerable to adversarial attacks, raising concerns about their robustness in real-world scenarios \cite{rabhi2024audio}\cite{farooq2025transferable}. These attacks typically introduce carefully crafted perturbations that are imperceptible to human listeners but can significantly alter model predictions. Both optimization-based and generative approaches have been shown to effectively degrade detection performance, even under black-box settings where model parameters are inaccessible. For instance, adversarial perturbations can drastically reduce detection accuracy while preserving perceptual quality, highlighting a critical security gap in existing ADD systems \cite{rabhi2024audio}. Moreover, such adversarial examples exhibit strong transferability across different model architectures, enabling attacks designed on one system to remain effective on unseen models \cite{farooq2025transferable}. Despite these advances, existing methods primarily rely on explicitly optimized perturbations with well-defined attack objectives. Consequently, the robustness of ADD systems under realistic and semantically preserving signal transformations remains insufficiently explored.

\subsection{Audio Watermarking}
Recent studies have explored deep learning-based audio watermarking techniques to improve security against synthetic speech misuse. WavMark \cite{chen2023wavmark} introduces a robust neural audio watermarking framework capable of embedding hidden messages while maintaining audio quality under common signal distortions. Similarly, \cite{liu2023detecting} proposed Timbre Watermarking, which embeds watermark information in the frequency domain to defend against unauthorized voice cloning attacks by preserving speaker-specific characteristics while enabling watermark extraction from synthesized speech. In addition, AudioSeal \cite{san2024proactive} presents a localized audio watermarking approach designed for efficient detection and generation-time verification of AI-generated speech. Recently, SilentCipher \cite{singh2024silentcipher} integrates psychoacoustic masking and pseudo-differentiable compression layers to achieve highly imperceptible yet robust watermarking, improving resistance against compression and signal distortions. Together, these audio watermarking models demonstrate the growing effectiveness of content authentication, ownership protection, and synthetic speech traceability. Nevertheless, the study of interaction between audio watermarking and ADD systems remains largely unexplored.

\section{Methodology}
\begin{figure}[t]
    \centering
\includegraphics[width=0.50\textwidth,height=6cm,keepaspectratio]{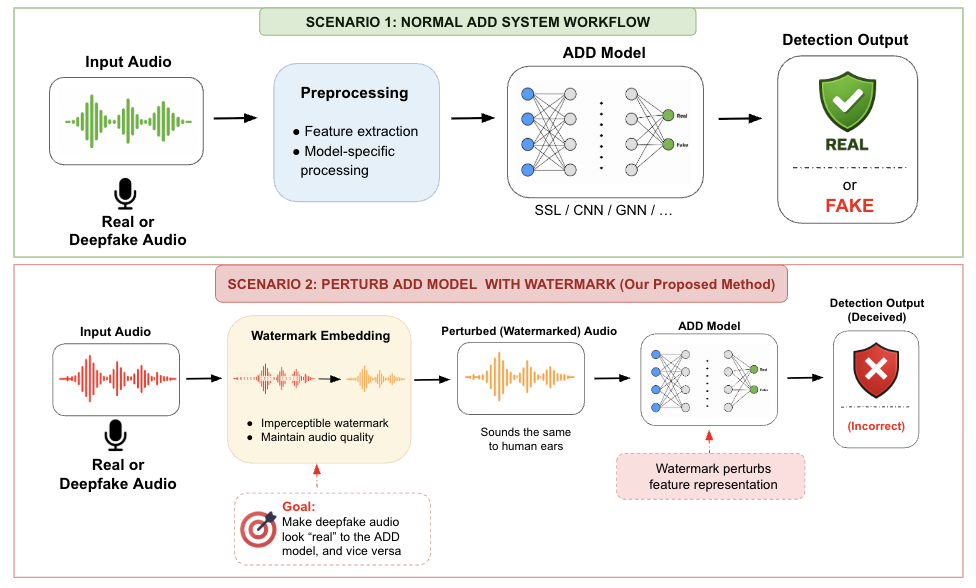}
    \caption{Overview of normal ADD inference and the proposed watermark-based perturbation framework.}
    \label{fig:overall}
\end{figure}
In this work, we investigate the robustness of ADD systems under watermark-induced perturbations. Furthermore, 
we also study how the signal modifications introduced by watermark embedding influence the feature representations and decision behaviour of ADD models.


\subsection{Watermark-based Evaluation Framework on ADD Models}
To systematically evaluate the robustness of ADD systems under structured signal perturbations, we propose a watermark-based evaluation framework on ADD models. 

\noindent \textbf{Audio Preprocessing.} In the following, we first denote $ \textbf{X} = [ \textbf{x}_{1}, \textbf{x}_{2}, \ldots,\textbf{x}_{S} ] \in \mathbb{R}^{S \times L}$ as the input audio clip of length $L$ (waveform samples) with $S$ different speakers. In this study, we set $L=16000$, corresponding to a one-second audio segment   sampled at 16 kHz. Each input waveform $\textbf{x}_S$ is then transformed into complex-valued spectral coefficients, via sliding window Fourier transform (SWFT), formulated as follows:
\begin{align} \label{eq:swft_transform_fwd}
\begin{split}
\mathcal{F}= \text{SWFT}_S(t,f) = 
\int_{-\infty}^{\infty} 
\mathbf{x_{\mathcal{S}}}(\tau) w(\tau-t)
e^{-i2\pi f\tau}d\tau,
\end{split}
\end{align}

\noindent
where $w(\tau-t)$ is the Hamming window function position at time $t$, and $e^{-i2\pi f\tau}$ is the Fourier basis used to extract the complex sinusoidal components at frequency $f$. 
The resulting spectral coefficients are then stacked into separated real $\mathrm{SWFT}^{\mathrm{real}}$ and imaginary components $\mathrm{SWFT}^{\mathrm{imag}}$ before being fed into the watermark embedding network, represented as a two channels representation corresponding to $
\mathrm{SWFT}_S = \mathrm{SWFT}_S^{\mathrm{real}} +
j\mathrm{SWFT}_S^{\mathrm{imag}} 
$. The watermarked waveform is reconstructed from the stacked spectral coefficients using the inverse SWFT, defined as follows:


\begin{equation}
\tilde{\bf x}_{S}
=
\frac{1}{C_{w}}
\int_{-\infty}^{\infty}
\int_{-\infty}^{\infty}
\tilde{\mathcal{F}}(t,f)
w(\tau-t)
e^{j2\pi f \tau}
\, df \, dt,
\label{eq:inverse_stft_continuous}
\end{equation}

\noindent
where \(C_w\) is a window-dependent normalization constant.

\noindent \textbf{Watermark Preprocessing.} 
The watermark message is represented as a binary vector $\mathbf{m}_{\mathrm{vec}} \in \{0,1\}^{K}$, where \(K\) denotes the watermark payload size in bits. Since the binary message has a lower dimensionality than the input waveform, a fully connected layer $\Gamma_{\mathrm{FC}}$ is used to project it into a latent representation with the same temporal length as the input waveform. This transformation can be expressed as: 
\begin{equation}
\mathbf{m}_{\mathrm{wave}}
=
\Gamma_{\mathrm{FC}}(\mathbf{m}_{\mathrm{vec}})
=
\mathbf{W}_{m}\mathbf{m}_{\mathrm{vec}}+\mathbf{b}_{m},
\end{equation}

\noindent
Here,
\(\mathbf{m}_{\mathrm{wave}} \in \mathbb{R}^{L}\) represents as the projected message representation, \(L\) denotes the number of waveform samples in one audio segment. 
where
\(\mathbf{W}_{m} \in \mathbb{R}^{L \times K}\) and 
\(\mathbf{b}_{m} \in \mathbb{R}^{L}\) are learnable weights and biases.
The projected message vector is further transformed using the same SWFT configuration applied to the input waveform, that can be written as: 
\begin{equation}
\mathbf{M}_{\mathrm{spec}}
=
\mathrm{SWFT}
(
\mathbf{m}_{\mathrm{wave}}
)
=
\mathrm{SWFT}
(
\Gamma_{\mathrm{FC}}
(
\mathbf{m}_{\mathrm{vec}}
)
),
\label{eq:message_stft}
\end{equation}

\noindent
where \(\
\mathbf{M}_{\mathrm{spec}}
\in
\mathbb{R}^{C \times F \times M},
\) with \(C=2\) corresponding to the real and imaginary channels of the SWFT representation. The $F$ and $M$ are the frequency and temporal dimensions, respectively. The transformed audio and watermark representations are concatenated along the channel dimension and fed into the invertible embedding network:

\begin{equation}
\mathbf{Z}^{0}
=
\mathrm{Concat}
\left(
\mathcal{F},
\mathbf{M}_{\mathrm{spec}}
\right),
\label{eq:concat}
\end{equation}

\noindent
where \(\mathbf{Z}^{0}\) denotes the joint representation containing both host-audio and watermark information. This concatenated representation is passed through an invertible neural network (INN), denoted by \(\mathcal{G}_{\theta}\), to produce the embedded spectral representation
\(\
\mathbf{Z}^{N}
=
\mathcal{G}_{\theta}
\left(
\mathbf{Z}^{0}
\right)
\)\
, where \(N\) denotes the number of invertible blocks and \(\theta\) represents the learnable parameters of the INN. The output
\(
\mathbf{Z}^{N}=
[\tilde{\mathcal{F}},\mathbf{R}
]
\) 
consists of watermarked spectrogram branch and a message branch, where only 
\(\tilde{\mathcal{F}}\) is retained. The final watermarked waveform is obtained using inverse SWFT,
\(
\bf\tilde{x}
=
\mathrm{iSWFT}
\left(
\tilde{\mathcal{F}}
\right)
\). 

\noindent \textbf{ADD Implementation.} In our work, we disregard the decoder branch since watermark extraction is not required. The watermarked audio $\tilde{\bf x}$ is then passed to a frozen ADD model $f_\phi^\ast(\cdot)$ to obtain the predicted probability of fake-audio:
\begin{align}
p_{\mathrm{fake}}
=
f_{\theta}(\tilde{\mathbf{x}}),
\end{align}
where $f_{\theta}(\cdot)$ denotes the ADD model parameterized by $\theta$, and 
$p_{\mathrm{fake}}\in[0,1]$ is the predicted probability that the input audio is classified as fake.

Unlike adversarial attacks that are intentionally optimized to fool a target model, watermark embedding introduces structured signal changes originally designed for content authentication. Since these modifications are part of a legitimate watermarking process, they preserve the semantic content and perceptual quality of the audio. Therefore, watermark-based perturbations provide a more realistic and practical evaluation setting than adversarial attacks, as they do not require access to model gradients or direct optimization against the target ADD model.

\subsection{Evaluation Metric}
We evaluate ADD performance using Equal Error Rate
(EER), which corresponds to the operating point where the
false acceptance rate equals the false rejection rate. Higher
EER indicates greater degradation of detection performance.

\subsection{Representation-level Evaluation}
To further investigate how watermarking affects feature representations, we extract intermediate embeddings from the ADD models for both $x$ and $\tilde{x}$. Let $z = g(x)$ and $\tilde{z} = g(\tilde{x})$ denote the embeddings extracted from the penultimate layer of an ADD model for the original and watermarked audio respectively, where $g(\cdot)$ represents the feature extraction function. The penultimate layer is selected because it captures high-level discriminative representations learned by the model before final classification. We analyze the impact of watermark-induced perturbations by comparing the distributions of $\{z\}$ and $\{\tilde{z}\}$ using multiple complementary metrics: \


\begin{table*}[!t]
\centering
\caption{EER (\%) of ADD Systems on Original (No WM) and Watermarked (WM) Audio Across Different Datasets}
\renewcommand{\arraystretch}{1.2}
\setlength{\tabcolsep}{5pt}

\resizebox{\textwidth}{!}{
\begin{tabular}{l cc | cc | cc | cc | cc}
\toprule
\multirow{2}{*}{\centering\textbf{ADD Models}} 
& \multicolumn{2}{c|}{\textbf{ASVspoof 2021 LA}} 
& \multicolumn{2}{c|}{\textbf{ASVspoof 2021 DF}} 
& \multicolumn{2}{c|}{\textbf{ASVspoof 2024}} 
& \multicolumn{2}{c|}{\textbf{FoR}}
& \multicolumn{2}{c}{\textbf{ITW}}\\

\cmidrule(lr){2-3} 
\cmidrule(lr){4-5} 
\cmidrule(lr){6-7}
\cmidrule(lr){8-9}
\cmidrule(lr){10-11}

& No WM & Proposed WM
& No WM & Proposed WM
& No WM & Proposed WM
& No WM & Proposed WM
& No WM & Proposed WM \\
\midrule

XLSR+SLS 
& 2.86 & 29.33 
& 1.91 & 33.90 
& 18.76 & 18.79 
& 5.07 & 6.38 
& 7.45 & 8.11 \\

TCM 
& 2.99 & 29.83 
& 2.14 & 34.22 
& 18.85 & 19.06 
& 10.68 & 11.20 
& 7.80 & 7.88 \\

Nes2Net 
& 2.17 & 29.70 
& 1.49 & 34.06 
& 22.05 & 21.67 
& 6.31 & 7.01 
& 7.75 & 8.55 \\
\midrule

AASIST 
& 11.46 & 30.83 
& 21.07 & 35.05 
& 35.53 & 35.59 
& 21.64 & 22.96 
& 43.00 & 43.86 \\

RawGAT-ST 
& 10.25 & 29.77 
& 23.26 & 36.50
& 40.29 & 40.41 
& 53.09 & 53.19 
& 52.53 & 53.79 \\
\midrule

RawNet2  
& 9.48 & 30.19 
& 22.38 & 36.44 
& 40.67 & 41.10 
& 48.54 & 47.49 
& 49.00 & 49.82 \\

RawTFNet 
& 5.04 & 29.84 
& 16.82 & 35.03 
& 44.73 & 41.37 
& 36.57 & 41.39 
& 38.72 & 46.00 \\

\bottomrule
\label{tab:results}
\end{tabular}}
\end{table*}

\noindent
{\textbf{Fréchet Distance (FD).}}
It has been widely used to evaluate distributional similarity in deep feature spaces, particularly in generative modeling and representation analysis \cite{heusel2017gans}. In this work, we use FD to quantify distributional differences between embeddings extracted from original and watermarked audio in the detector feature space. Larger FD values indicate stronger feature-space distortion induced by watermark perturbations.



\noindent
\textbf{Cosine and L2 Distance.} 
We also measure similarity between paired embeddings $(z, \tilde{z})$ using cosine similarity and L2 distance:
Cosine similarity captures angular deviations in the feature space, while L2 distance reflects absolute magnitude differences, with larger values indicating stronger perturbations.

\subsection{Models Selection}
To evaluate the robustness of watermark-induced perturbations across different architectures, we select representative ADD models from three major categories: self-supervised learning (SSL), graph neural network (GNN), and convolutional neural network (CNN) based systems. The model selection is guided by Speech DF Arena \cite{dowerah2026speech}, focusing on methods that (i) achieve competitive benchmark performance, (ii) are published in leading venues, and (iii) provide publicly available pretrained models for reproducibility.   All selected ADD models
are trained on the ASVspoof 2019 dataset. For watermarking model, we use WavMark \cite{chen2023wavmark} as the watermarking framework. All audio samples were resampled to 16 kHz, and watermark embedding was performed using the pretrained WavMark model\footnote{\url{https://github.com/wavmark/wavmark}} with default 16 watermarking bits. WavMark achieves low-distortion watermarking with an SNR of 38.55 dB and a PESQ score of 4.3 while preserving semantic audio content, making it suitable for controlled robustness evaluation.

\subsection{Evaluation Datasets}
We evaluate watermark-induced robustness across five benchmark audio deepfake detection datasets: ASVspoof 2021 LA and DF \cite{yamagishi2021asvspoof}, ASVspoof 2024 \cite{wang2024asvspoof}, In-the-Wild (ITW) \cite{muller2022does}, and Fake-or-Real (FoR) \cite{reimao2019dataset}. ASVspoof 2021 LA and DF primarily contain relatively controlled and homogeneous recording conditions while ASVspoof 2024 introduces more realistic acoustic variability, codec diversity, and advanced spoof generation conditions. The ITW dataset contains real-world deepfake audio collected from online and social media sources under diverse recording environments and unseen synthesis pipelines. Similarly, FoR includes both high-fidelity and rerecorded telephone-like conditions designed to evaluate robustness under channel distortions. These datasets collectively enable analysis of watermark-induced perturbations across both controlled benchmark settings and realistic real-world conditions.

\section{Results and Analysis}
\subsection{Impact of Watermarking on ADD models} 
Table \ref{tab:results} summarises the equal error rate (EER) of ADD models before and after watermarking. A key observation is that watermarking causes severe performance degradation on the ASVspoof 2021 LA and DF datasets across all evaluated architectures, with EER increases up to 36.50\%. In contrast, watermarking has lesser impact on ASVspoof 2024, FoR and ITW, where most models exhibit marginal EER changes. Compared with ASVspoof 2021 LA and DF, these datasets contain substantially greater variability in recording conditions, transmission channels, and spoof generation pipelines. In contrast, ASVspoof 2021 LA and DF are relatively clean and controlled benchmark datasets with more homogeneous spoofing characteristics. As a result, watermark perturbations may more strongly affect the subtle patterns used to identify fake audio in the relatively homogeneous ASVspoof 2021 LA and DF datasets, leading to larger performance degradation. The increased variability in ASVspoof 2024, FoR and ITW likely reduces dependence on such fragile cues, resulting in more stable performance under watermarking.

\begin{table*}[!t]
\centering
\caption{Representation-Level Embedding Shifts Under Watermark Perturbation}
\renewcommand{\arraystretch}{1.2}
\setlength{\tabcolsep}{4pt}

\begin{tabular}{c| l | cc| cc| cc | cc | cc}
\toprule

\multirow{3}{*}{} 
&\multicolumn{1}{c|}{\multirow{3}{*}{\textbf{Metrics}}}& \multicolumn{6}{c|}{\textbf{ASVspoof}} 
& \multicolumn{2}{c|}{\textbf{FoR}} 
& \multicolumn{2}{c}{\textbf{ITW}} \\

& 
& \multicolumn{2}{c|}{\textbf{2021 LA}} 
& \multicolumn{2}{c|}{\textbf{2021 DF}} 
& \multicolumn{2}{c|}{\textbf{2024}} 
& \multicolumn{2}{c|}{} 
& \multicolumn{2}{c}{} \\

\cmidrule(lr){3-4}
\cmidrule(lr){5-6}
\cmidrule(lr){7-8}
\cmidrule(lr){9-10}
\cmidrule(lr){11-12}

& 
& Spoof & Bonafide 
& Spoof & Bonafide 
& Spoof & Bonafide 
& Spoof & Bonafide 
& Spoof & Bonafide \\

\midrule

\multirow{3}{*}{\centering{\textbf{XLSR+SLS}}}
& Fréchet Distance & 419.10 & 37.52 & 498.55 & 12.25 & 2.76 & 1.92 & 55.85 & 160.65 & 81.73 & 271.14 \\
& Mean Cosine Distance & 0.30 & 0.06 & 0.30 & 0.04 & 0.01 & 0.01 & 0.14 & 0.13 & 0.22 & 0.20 \\
& Mean L2 Distance & 22.09 & 10.55 & 23.84 & 9.10 & 4.77 & 4.13 & 18.25 & 17.94 & 21.63 & 21.40 \\
\midrule

\multirow{3}{*}{\centering{\textbf{TCM}}}
& Fréchet Distance & 103.90 & 9.63 & 91.25 & 2.78 & 1.55 & 0.48 & 7.85 & 19.35 & 3.70 & 0.67 \\
& Mean Cosine Distance & 0.66 & 0.13 & 0.66 & 0.07 & 0.04 & 0.01 & 0.12 & 0.16 & 0.12 & 0.03 \\
& Mean L2 Distance & 10.67 & 4.41 & 11.42 & 3.58 & 2.89 & 1.55 & 4.51 & 5.23 & 5.02 & 2.01 \\
\midrule

\multirow{3}{*}{\centering{\textbf{Nes2Net}}}
& Fréchet Distance & 965.67 & 128.98 & 1089.69 & 35.92 & 4.51 & 1.05 & 99.41 & 219.01 & 5.94 & 7.79 \\
& Mean Cosine Distance & 0.31 & 0.14 & 0.38 & 0.09 & 0.03 & 0.02 & 0.15 & 0.15 & 0.05 & 0.02 \\
& Mean L2 Distance & 32.50 & 7.95 & 34.89 & 5.11 & 8.77 & 2.10 & 14.06 & 11.56 & 9.45 & 2.27 \\
\midrule

\multirow{3}{*}{\centering{\textbf{AASIST}}}
& Fréchet Distance & 1.12 & 26.28 & 2.16 & 58.04 & 0.22 & 0.10 & 0.09 & 0.06 & 0.04 & 0.06 \\
& Mean Cosine Distance & 0.06 & 0.15 & 0.15 & 0.32 & 0.00 & 0.00 & 0.00 & 0.00 & 0.01 & 0.02 \\
& Mean L2 Distance & 2.81 & 3.94 & 5.29 & 8.11 & 0.78 & 0.68 & 0.71 & 0.49 & 0.58 & 0.76 \\
\midrule

\multirow{3}{*}{\centering{\textbf{RawGAT-ST}}}
& Fréchet Distance & 0.85 & 18.06 & 0.27 & 12.55 & 0.10 & 0.02 & 0.21 & 0.44 & 0.00 & 0.03 \\
& Mean Cosine Distance & 0.19 & 0.69 & 0.17 & 0.60 & 0.03 & 0.02 & 0.25 & 0.08 & 0.10 & 0.09 \\
& Mean L2 Distance & 3.17 & 4.66 & 2.78 & 4.62 & 1.14 & 0.78 & 4.52 & 2.35 & 0.54 & 0.62 \\
\midrule

\multirow{3}{*}{\centering{\textbf{RawNet2}}}
& Fréchet Distance & 2.65 & 161.65 & 0.31 & 85.36 & 0.00 & 0.00 & 0.00 & 0.00 & 0.04 & 0.37 \\
& Mean Cosine Distance & 0.15 & 0.72 & 0.08 & 0.57 & 0.00 & 0.00 & 0.00 & 0.00 & 0.05 & 0.09 \\
& Mean L2 Distance & 2.99 & 11.94 & 2.12 & 10.19 & 0.09 & 0.10 & 0.07 & 0.08 & 0.63 & 1.04 \\
\midrule

\multirow{3}{*}{\centering{\textbf{RawTFNet}}}
& Fréchet Distance & 12.39 & 3.86 & 4.42 & 0.11 & 0.12 & 0.08 & 0.01 & 0.02 & 0.33 & 0.15 \\
& Mean Cosine Distance & 0.51 & 0.32 & 0.44 & 0.12 & 0.01 & 0.01 & 0.00 & 0.00 & 0.05 & 0.02 \\
& Mean L2 Distance & 3.42 & 2.32 & 2.95 & 1.28 & 0.43 & 0.38 & 0.27 & 0.33 & 0.78 & 0.56 \\

\bottomrule
\end{tabular}
\label{tab:embedding}
\end{table*}
\subsection{Representation-Level Analysis}
To further analyse the impact of watermarking, Table \ref{tab:embedding} provides a deeper representation-space interpretation of the degradation trends observed in Table \ref{tab:results} by quantifying how watermark perturbation alters the embedding representation of ADD systems.

The first observation is that conventional SSL-based models exhibit substantial embedding distribution shifts after watermark insertion. XLSR+SLS consistently shows the largest Fréchet distances across nearly all datasets, particularly on ASVspoof 2021 DF where the FD reaches 498.55 for spoof samples. Similar behavior is also observed on ASVspoof 2021 LA with FD values of 419.10 for spoof embeddings. These extremely large distributional deviations directly align with the severe EER degradation reported in Table \ref{tab:results}, where XLSR+SLS increases from 2.86\% to 29.33\% on LA and from 1.91\% to 33.90\% on DF under watermark perturbation. The same trend is reflected in TCM and Nes2Net, both of which demonstrate large representation drifts accompanied by substantial EER inflation. For example, Nes2Net produces FD values of 965.67 on LA and 1089.69 on DF for spoof embeddings, while its EER rises from 2.17\% to 29.70\% and from 1.49\% to 34.06\%, respectively. These results suggest that watermark perturbation significantly distorts the learned latent manifolds of these architectures, causing spoof and bonafide distributions to move away from the original decision regions.

\begin{figure}[t]
    \centering
    \begin{subfigure}{0.48\columnwidth}
        \centering
        \includegraphics[width=\linewidth]{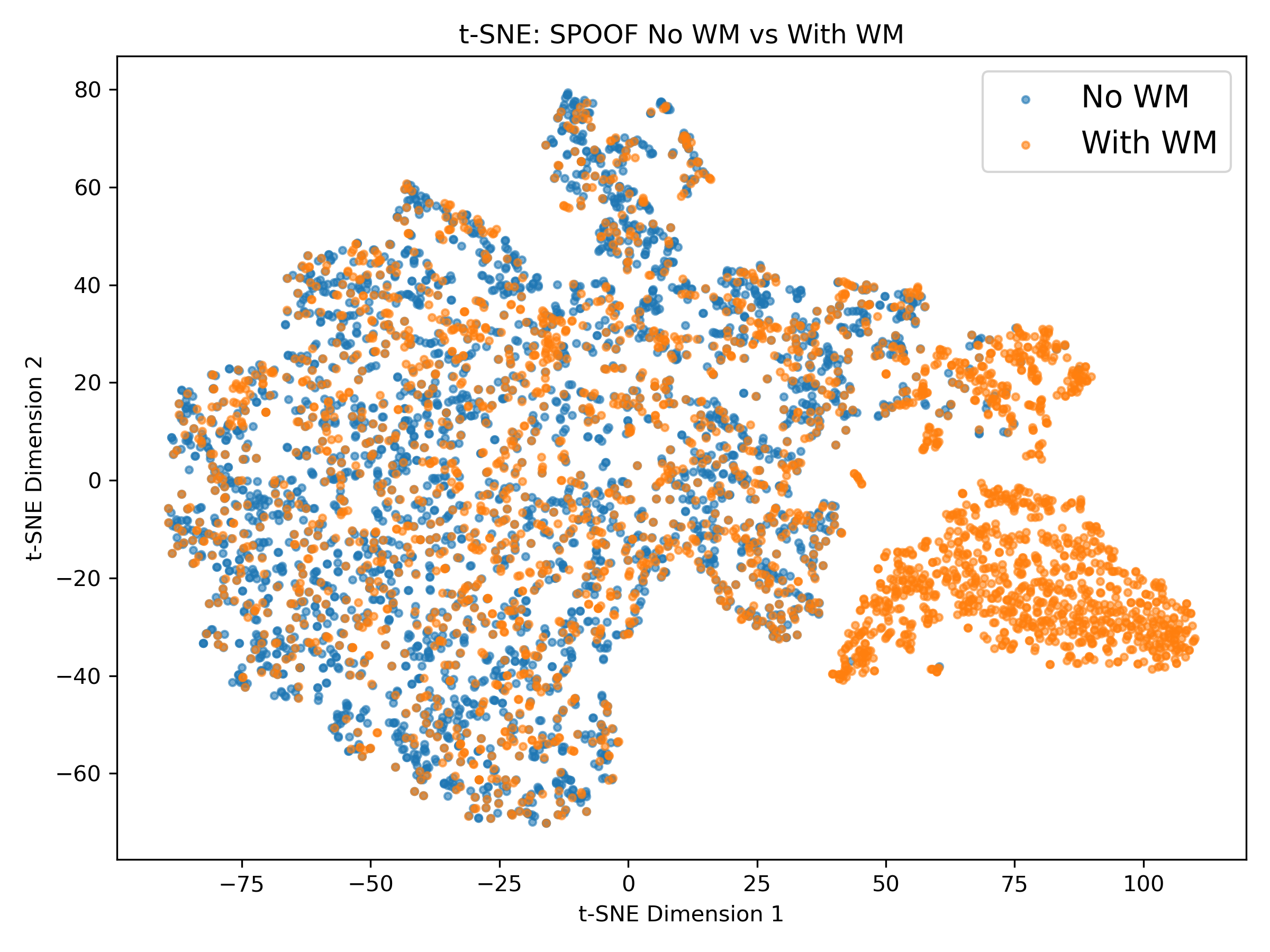}
        \caption{XLSR+SLS — Spoof}
    \end{subfigure}
    \hfill
    \begin{subfigure}{0.48\columnwidth}
        \centering
        \includegraphics[width=\linewidth]{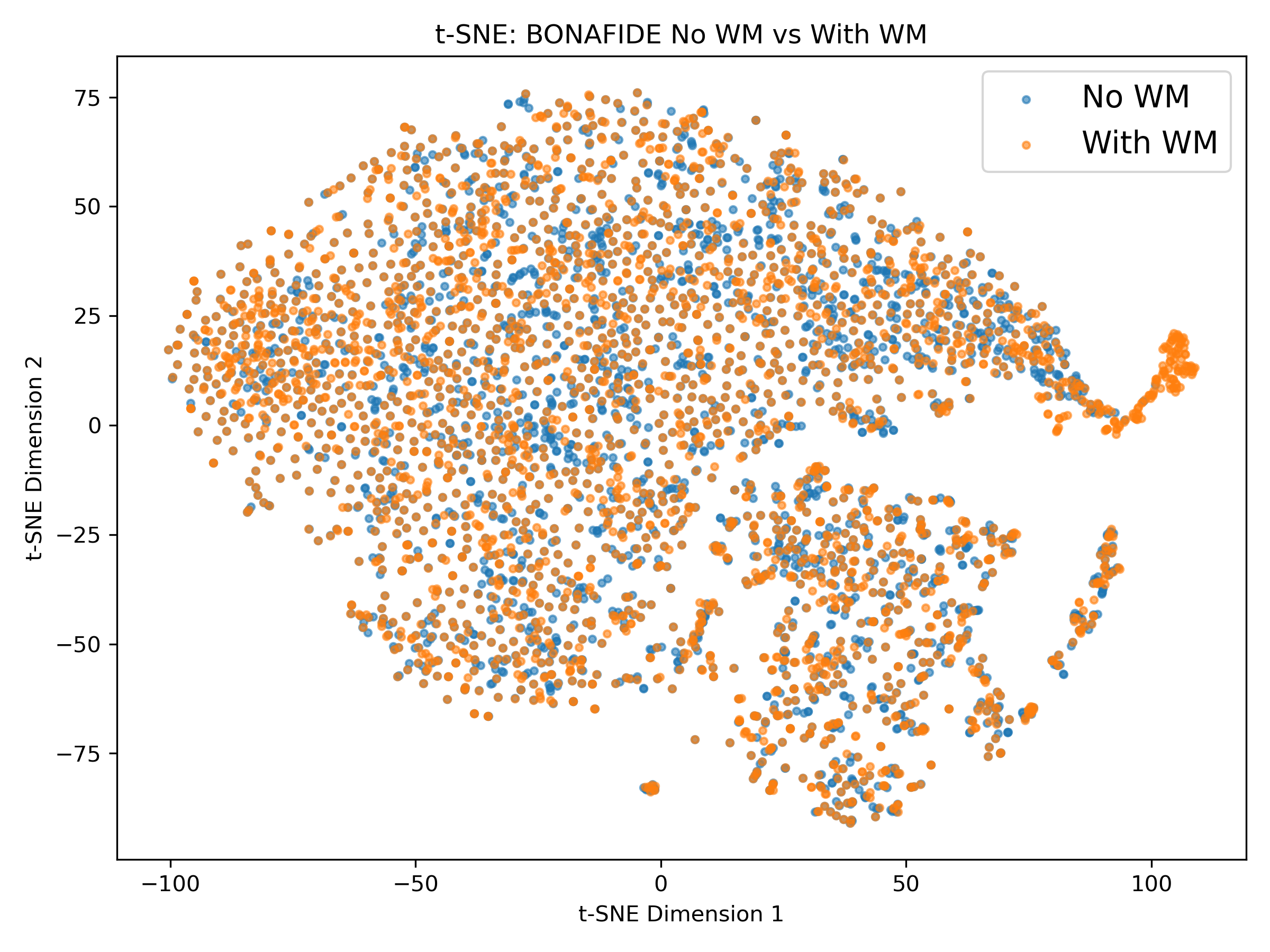}
        \caption{XLSR+SLS — Bonafide}
    \end{subfigure}

    \vspace{0.3em}

    \begin{subfigure}{0.48\columnwidth}
        \centering
        \includegraphics[width=\linewidth]{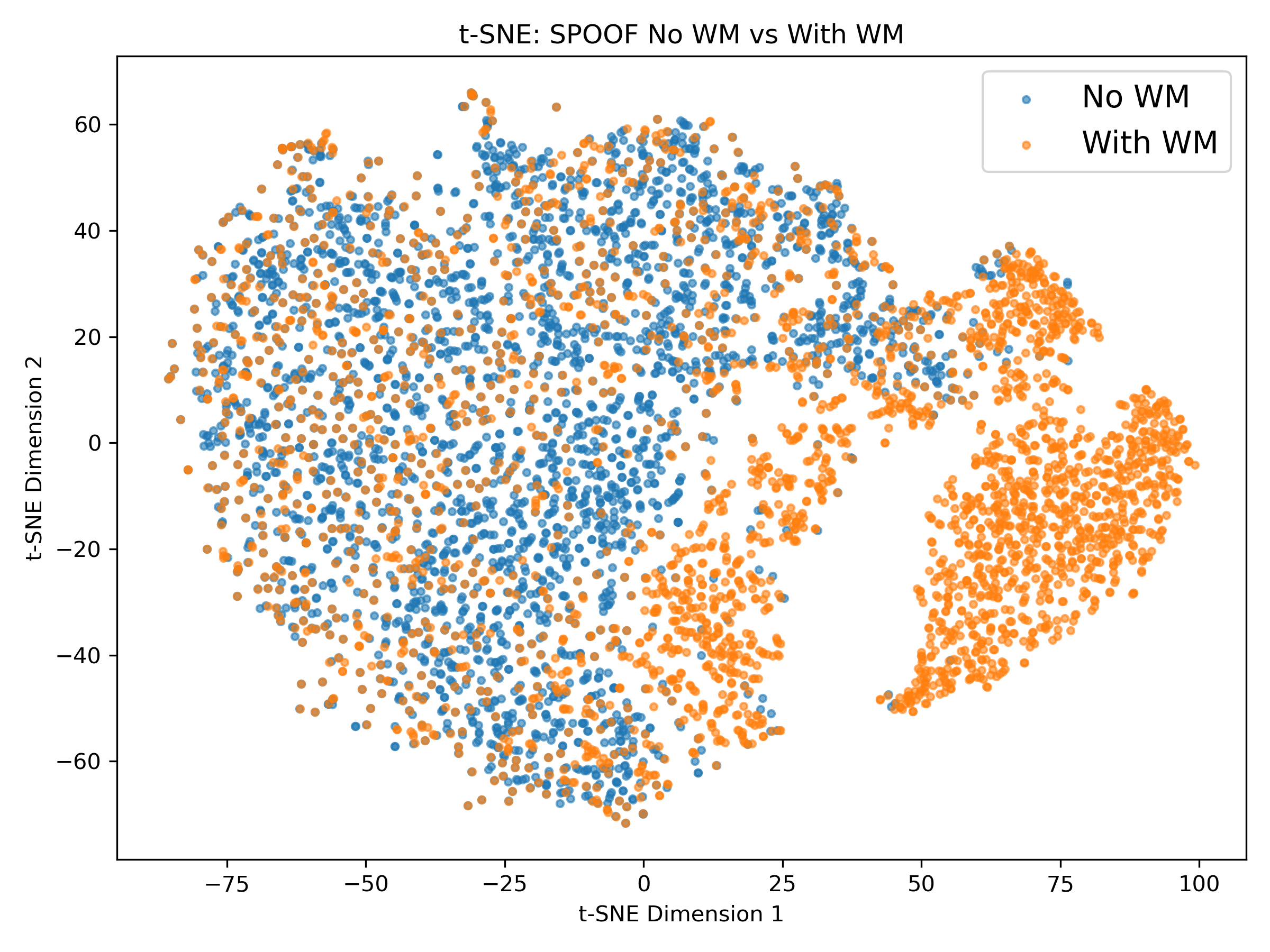}
        \caption{TCM — Spoof}
    \end{subfigure}
    \hfill
    \begin{subfigure}{0.48\columnwidth}
        \centering
        \includegraphics[width=\linewidth]{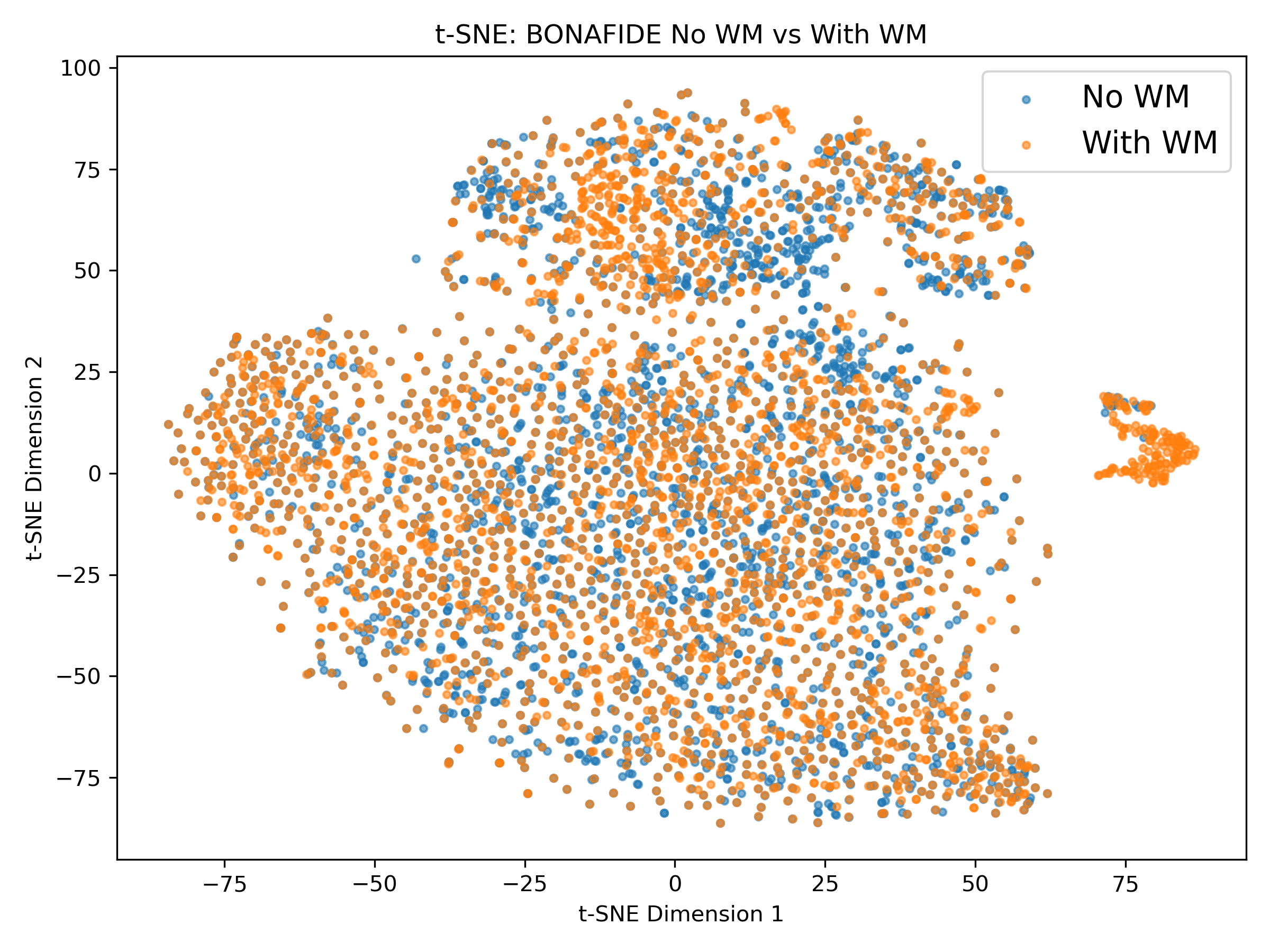}
        \caption{TCM — Bonafide}
    \end{subfigure}

    \vspace{0.3em}

    \begin{subfigure}{0.48\columnwidth}
        \centering
        \includegraphics[width=\linewidth]{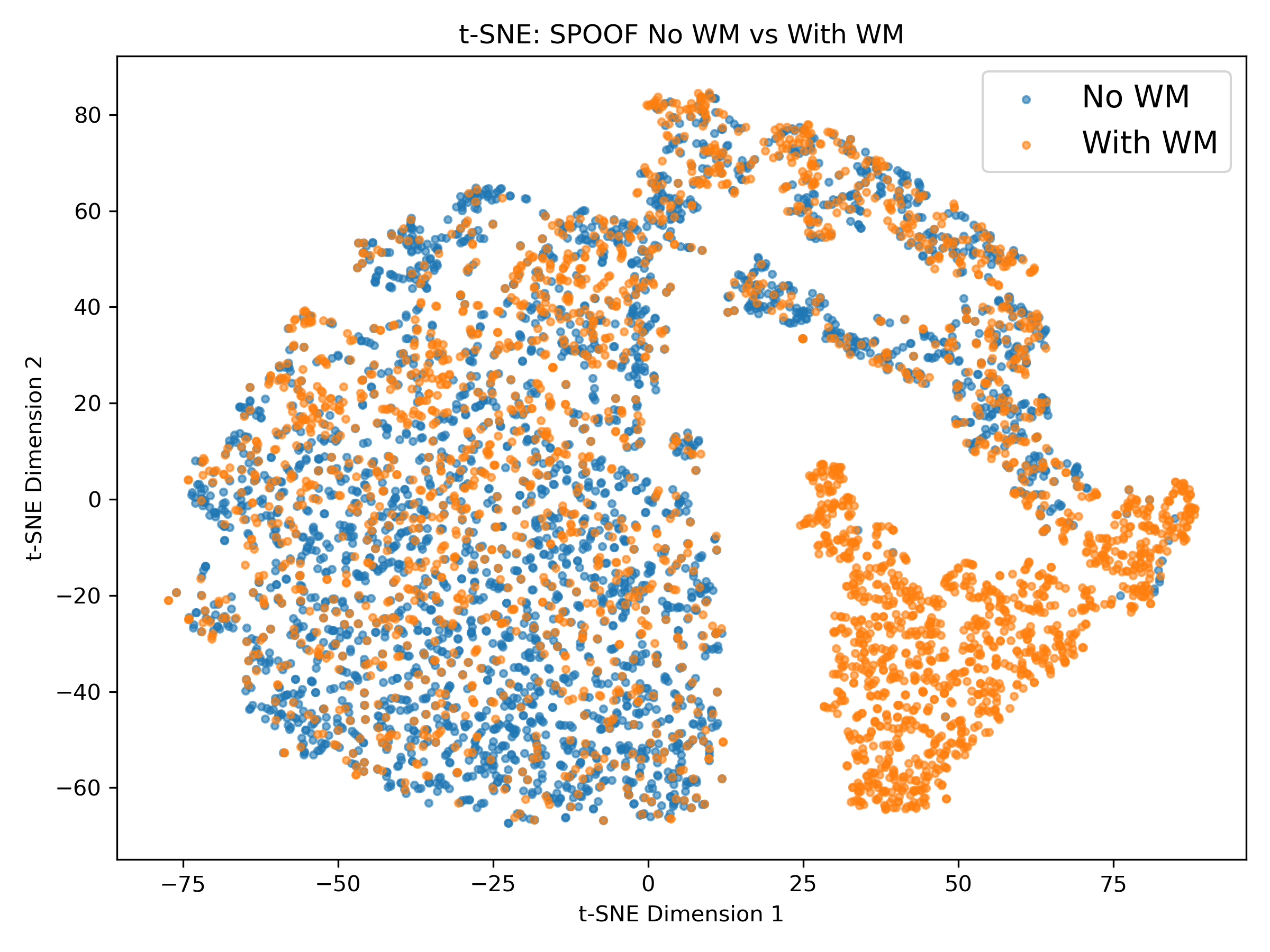}
        \caption{Nes2Net — Spoof}
    \end{subfigure}
    \hfill
    \begin{subfigure}{0.48\columnwidth}
        \centering
        \includegraphics[width=\linewidth]{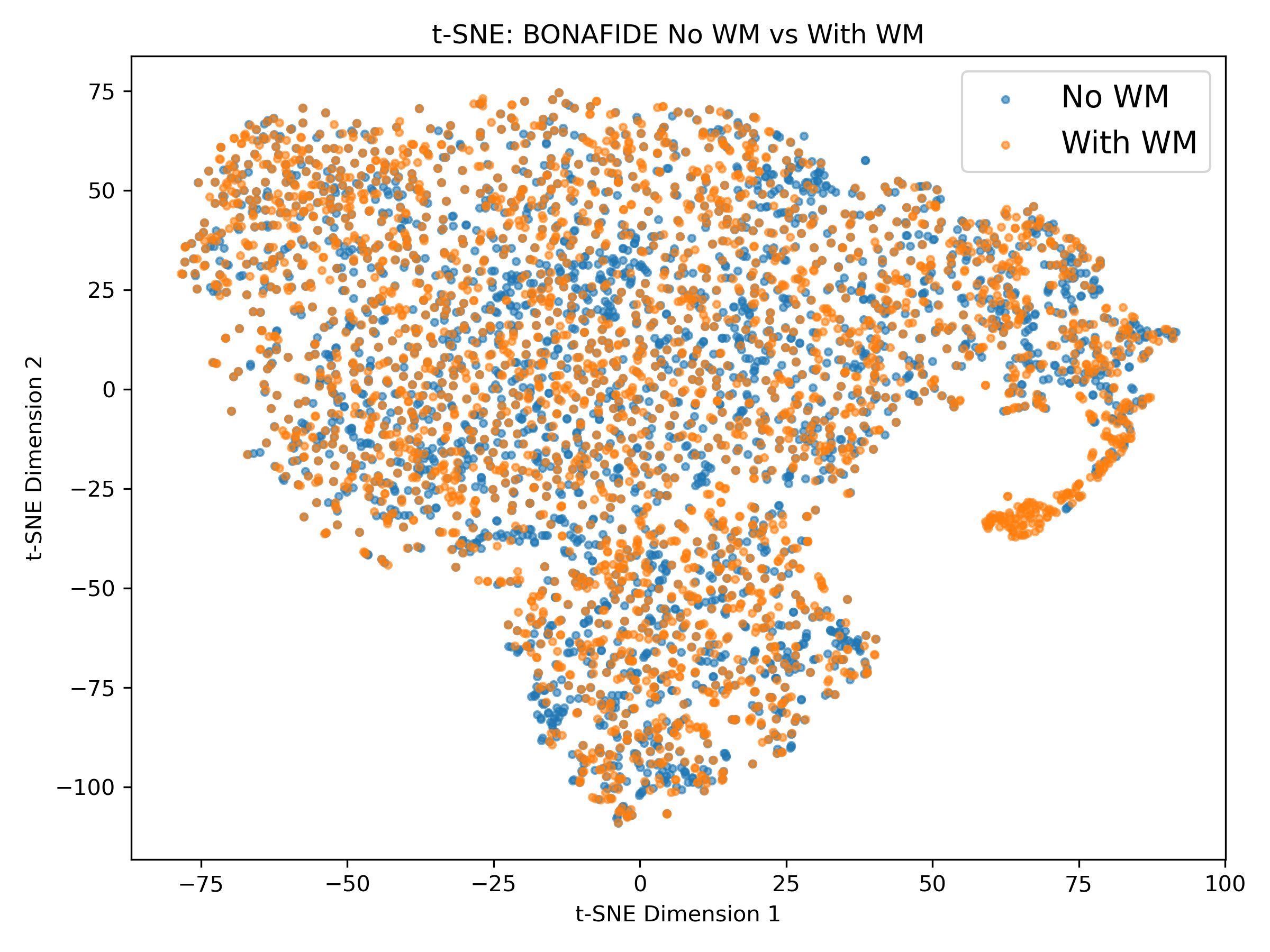}
        \caption{Nes2Net — Bonafide}
    \end{subfigure}

    \caption{
    t-SNE visualization of embedding distributions before (blue) and after (orange) watermark perturbation on the ASVspoof 2021 LA dataset. Larger distribution shifts between the embeddings suggest that watermarking alters the learned feature space and affects ADD decision boundaries.}
    \label{fig:tsne}
\end{figure}

Another key finding is that SSL-based models, particularly XLSR+SLS, TCM and Nes2Net exhibit substantially larger embedding perturbations for fake (spoof) samples than for real (bonafide) samples across most datasets. This suggests that spoof representations exhibit greater sensitivity to watermark-induced perturbations, resulting in larger representation-space shifts that are consistently reflected by higher spoof-side FD, cosine and L2 differences. We then further validate this observation through the t-SNE visualizations shown in Fig. \ref{fig:tsne}, where spoof embeddings exhibit clear cluster migration and separation between the original and watermarked samples. In particular, XLSR+SLS, TCM, and Nes2Net show obvious watermark-induced spoof clusters, indicating substantial deformation of the latent spoof representation space. In contrast, the bonafide embeddings remain comparatively more stable, implying that watermark perturbations disrupt spoof-specific representations more strongly than bonafide representations.

The GNN-based architectures such as AASIST and RawGat-ST exhibit relatively smaller embedding shifts despite still experiencing EER degradation. They maintain comparatively low FD, cosine and L2 distances across nearly all datasets. For instance, AASIST produces FD values as low as 1.12 on LA spoof and 2.16 on DF spoof, while RawGAT-ST remains below 1.0 FD. The corresponding cosine distance is also significantly smaller than those observed in SSL-based systems. This behavior indicates that these models preserve a more stable latent representation structure under watermark perturbation. Nevertheless, Table \ref{tab:results} shows that their EER still increases substantially under watermarking, implying that even relatively small feature-space shifts can critically affect the classifier decision boundary in anti-spoofing tasks.

Interestingly, CNN-based models such as RawNet2 and RawTFNet demonstrate a moderate behavior. Although their embedding shifts are lower than those of SSL-based systems, they still exhibit noticeable displacement in some scenarios, particularly on dataset ASVspoof 2021 LA and DF. RawNet2 shows elevated FD values on LA bonafide embeddings (161.65), whereas RawTFNet exhibits moderate but consistent shifts across all metrics. Correspondingly, their EER degradation is also moderate relative to XLSR+SLS and Nes2Net. This consistency between representation displacement and detection performance further strengthens the hypothesis that watermark perturbation disrupts the stability of latent feature distributions, ultimately degrading spoof discrimination capability.

Another notable observation is the dataset dependency of the perturbation effect. The embedding shifts are substantially larger on ASVspoof 2021 LA and DF compared with ASVspoof 2024, FoR and ITW. It could be due to the differences in dataset construction and acoustic variability. ASVspoof 2021 is largely derived from the VCTK corpus, which was recorded in a studio-quality semi-anechoic environment and therefore contains relatively clean and controlled speech conditions. On the other hand, ASVspoof 2024 introduces crowdsourced speech collected from a vastly larger number of speakers under diverse acoustic conditions, together with stronger spoofing and adversarial attacks. Similarly, FoR includes re-recorded and telephone-like degraded conditions, while the ITW dataset is to evaluate real-world generalization beyond controlled ASVspoof benchmarks. Compared with the relatively clean ASVspoof 2021 conditions, these datasets already contain substantial acoustic variability, codec distortion, replay effects and environmental noise. As a result, the spoof-related cues are inherently less sensitive to additional perturbations, making the embedding space comparatively more robust against watermark-induced distortion. Consequently, ASVspoof 2024, FoR and ITW generally exhibit smaller FD and cosine shifts across most of the ADD models.


Overall, Table \ref{tab:results} and Table \ref{tab:embedding} show that watermark perturbation alters the latent representation space of ADD systems, particularly on ASVspoof 2021 LA and DF. Rather than behaving as simple additive noise, the watermark introduces structured distortions that disrupt the learned embedding manifold, reduce spoof–bonafide separability, and affect the downstream decision boundary. These findings highlight the vulnerability of current ADD systems to representation-level perturbations introduced by embedded watermark signals.

\section{Conclusion}
In this work, we studied the robustness of ADD systems under audio watermarking. Results show that watermark embedding can significantly alter learned feature representations, particularly in SSL-based models, leading to degraded spoof detection performance. These findings reveal that watermarking techniques designed for content authentication may unintentionally weaken existing ADD systems. Unlike conventional adversarial attacks, watermark perturbations require neither gradient access nor explicit optimization, making them more practical in real-world media pipelines. Overall, these findings highlight the need for more robust ADD systems under realistic watermarking transformations.


\bibliographystyle{IEEEtran}
\bibliography{strings}

\end{document}